# Automated Vehicle Location (AVL) Using Global Positioning System (GPS)

Victor Dutta, R. Bera, Sourav Dhar, Jaydeep Chakravorty, Nishant Bagehel
Sikkim Manipal Institute of Technology, Majhitar, Rangpo, East Sikkim-737132 (INDIA)
vic.excuse@gmail.com, rnbera@gmail.com

*Abstract*—Generally vehicular networks are considered to contain two types of nodes; vehicles and roadside stations. Vehicular communications is usually developed as a part of Intelligent Transport Systems (ITS) that seeks to achieve safety and productivity through intelligent transportation which integrates communication between mobile and fixed nodes. The majority of today's automated vehicle location (AVL) systems use Global Positioning System (GPS) technology. Each satellite has its individual code (waveform) which contains the required data to calculate the position of the receiver. GPS receivers have been miniaturized to just a few integrated circuits and so are becoming very economical and can easily be fitted inside vehicles. Low-cost Differential GPS (DGPS) receivers, which have a positioning accuracy of approximate 2–3 m, have become available. It is now possible to perform AVL down to specific roadway lanes. In this paper, a vehicle-lane-determining system is described, consisting of an onboard DGPS receiver that is connected with a wireless communications channel, a unique lane-level digital roadway database, a developed lane-matching algorithm, and a real-time vehicle location display. Surveyed map data based on more than 100 000 s, it has correctly determined the lane 97% of the time.

## I. INTRODUCTION

GLOBAL positioning system (GPS) operates on the principal of relative calculation of position by demodulating the electro-magnetic signals sent by the GPS satellites. The signals contain information about the time when the message was sent. Each satellite has its individual code (waveform) which provides identity to the satellite and contains the information necessary for calculating the position of the receiver. The vehicular network uses GPS to find its position and communicate. The nodes ie; vehicles and roadside stations are Dedicated Short Range Communications (DSRC) devices. DSRC works in 5.9GHz band with bandwidth of 75 MHz and approximate range of 1000m.[1] The network should support both private data communications and public (mainly safety) communications.

GPS is helpful in getting the perfect timing on the clocks on earth. The clocks on the GPS satellites are atomic clocks which give very precise timing. These atomic clocks use the time period of oscillation of a specific atom as a standard to measure the time. The clocks on earth can be matched to the time of these clocks to get accuracy. This can also be used to mark the time simultaneously at different places. For example if in the global share market a transaction takes place in INDIA at a certain instant the trading I registered at all the share markets in the world without any regional time delay.

Tracking is another important use of GPS positioning. This paper deals with this use of GPS. The vehicles can be fitted with GPS receivers which can track their positions and hence we can know where a vehicle is at any instant of time. By this way we can know the traffic in any lane and we can manage the fleet of cars.

GPS basically was designed for navigation and today it is being used in all sorts of operations. It is being used in army, fishing, farming and timing too.

## II. TRIANGULATION

The receiver uses a mathematical process known as triangulation to calculate its position with respect to the satellites. This process can be said to be trilateration rather triangulation because no angles are involved. It is a process of finding the relative position of an object using geometry of a triangle. The signal sent by the satellite contains information about the time when the information was sent (Ts). The time at which the receiver obtains the signal is also noted down (Tr). So the total time taken by the signal to reach the receiver is (Tr-Ts). We know the EM waves travel with the speed of light so the DISTANCE TRAVELED=SPEED X TIME TAKEN. So the distance between the satellite and the receiver is D=(Tr-Ts)c. Now we assume a sphere with a radius equal to D and consider the receiver to be somewhere on the surface of the sphere. This limits the position of the receiver to only few points in the vast universe.

If x1, y1, z1 be the coordinates of the satellite in the universe and D1 be its distance from the receiver then equation of the sphere is,

$$(x-x_i)^2 + (y-y_i)^2 + (z-z_i)^2 = ((tr_i + b - t_i)c)^2 \quad \ldots(i)$$

for i=1,2,3,4.
Where,

$x_i$, $y_i$ and $z_i$ are the coordinates of the four satellite positions.

$tr_i$ -the receiving time of the signal. $t_i$ - the time at which the signal was sent. c-the speed of light

Similarly by considering the sphere for the next satellite we can find out the intersections of the two spheres. The points that we get lie on a circle. (Fig.1)

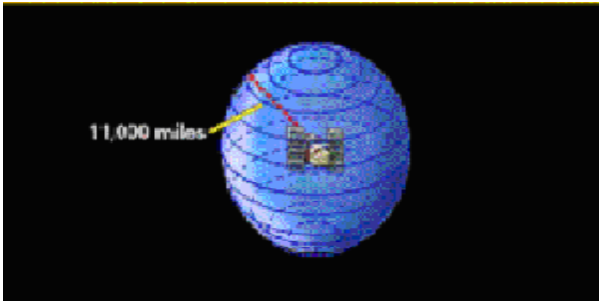

Fig.1: Sphere assumption of a satellite

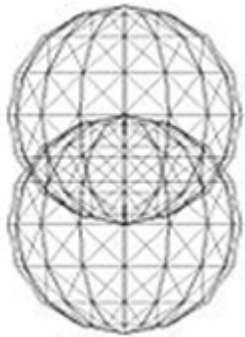

Fig 2: Sphere intersection

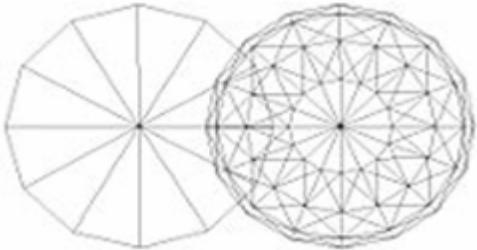

Fig.3: Intersection of circle & sphere

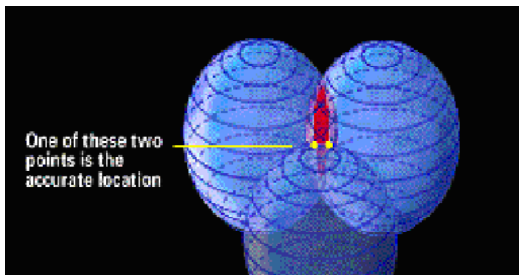

Fig.4: Accurate location of receiver

So now we are confined only within circle. This step is followed by again taking the intersection of this circle and the sphere due to the next satellite. The process is known as triangulation. (Fig.2).The intersection of the third sphere and the circle gives 2 or 1 points where the receiver can be placed.

The exact position can be calculated by solving the equation for three spheres and equating them to get the points of intersection. (Fig.3). From equation (i), we get two values for the position of the receiver. One of the values is absurd as it is far away from the surface of the earth which is not possible. So we discard one of the values and the other value is the exact position of the receiver (Fig.4).

### III. CALCULATION OF THE TIME BIAS

After considering the three spheres for calculation we can also use a fourth sphere for more accurate position finding. Due to large distance between the receiver and the satellite many errors creep in. For example due to ionospheric noise the signal may take longer to reach the receiver. This leads to errors in calculation of position. The fourth sphere may not intersect exactly at one of the points. To take care of this error the concept of pseudo ranges is taken into consideration.

The distance between the probable point calculated above and the fourth satellite is calculated and termed as pseudo range(P4). Also the radius of fourth sphere as calculated from the signal with error is taken (r4). Let $D_a$=(r4-P4) be the distance between the computed GPS receiver position and surface of sphere corresponding to the fourth satellite. Thus the quotient b=$D_a$/c provides an estimate of:(correct time)-(time indicated by the receivers on board clock).The receivers clock can be advanced if b is positive and delayed if b is negative.

### IV. CALCULATING THE COORDINATES OF SATELLITES

The earth is divided into 180 latitudes and an equivalent number of longitudes. The equator is the 0 degree latitude and the poles are the 0 degree longitude. The X and Y coordinates of the satellite will be the same on earth surface as that of a point right under the satellite. Considering the point of intersection of 0 degree latitude and longitude as the origin we can represent the position of the satellite.

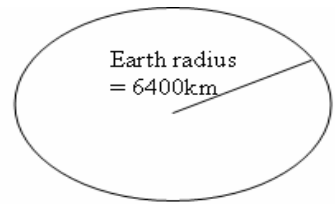

Fig.5: An example of calculation

From Fig.5,

The circumference of earth = $2\pi R$ =40212.38km

Now the length of the quarter circumference = 10053.09km

This length contains 90 latitudes (from equator to the pole). So the distance covered on moving one latitude up or down= $10053.09/90$ =117.70km

Similarly for the longitudes we have 90 longitudes spread across the semi circumference of the earth. So on moving across one longitude we cover a distance $= 20106.19/90 = 223.40$ km.

So we can calculate the X & Y components by converting the degree into kilometers. The source of error can be, we have considered the earth to be completely spherical whereas it is oval in shape.

For example a satellite with coordinates

Latitude=49.6 degree, Longitude=40.6 degree, Height =27500km

The interpretation in terms of Cartesian coordinates would be

Along X axis= $49.6 \times 117.70 km$ =5540.32km

Along Y axis= $40.6 \times 223.40 km$ =9070.12km

Along Z axis=27500km

## V. GOLD CODES

The signal sent by the satellites is an electro-magnetic wave containing certain information. This signal is known as gold code or navigation message. Each GPS satellite continuously broadcasts a **Navigation Message** at 50 bit/s giving the time-of-week, GPS week number and satellite health information (all transmitted in the first part of the message), an *EPHEMERIS* (transmitted in the second part of the message) and an *ALMANAC* (later part of the message). The messages are sent in frames, each taking 30 seconds to transmit 1500 bits. The first 6 seconds of every frame contains data describing the satellite clock and its relationship to GPS time. The next 12 seconds contain the **ephemeris** data, giving the satellite's own precise orbit. The ephemeris is updated every 2 hours and is generally valid for 4 hours, with provisions for updates every 6 hours or longer in non-nominal conditions. The time needed to acquire the ephemeris is becoming a significant element of the delay to first position fix, because, as the hardware becomes more capable, the time to lock onto the satellite signals shrinks, but the ephemeris data requires 30 seconds (worst case) before it is received, due to the low data transmission rate.

The **almanac** consists of coarse orbit and status information for each satellite in the constellation, an ionospheric model, and information to relate GPS derived time to Coordinated Universal Time (UTC). A new part of the almanac is received for the last 12 seconds in each 30 second frame. Each frame contains 1/25th of the almanac, so 12.5 minutes are required to receive the entire almanac from a single satellite. The almanac serves several purposes. The first is to assist in the acquisition of satellites at power-up by allowing the receiver to generate a list of visible satellites based on stored position and time, while an ephemeris from each satellite is needed to compute position fixes using that satellite. Finally, the almanac allows a single frequency receiver to correct for ionospheric error by using a global ionospheric model. An important thing to note about navigation data is that each satellite transmits only its own *ephemeris*, but transmits an *almanac* for all satellites.

Each satellite transmits its navigation message with at least two distinct spread spectrum codes: the Coarse / Acquisition (C/A) code, which is freely available to the public, and the Precise (P) code (Fig.6), which is usually encrypted and reserved for military applications. The C/A code is a 1023 length Gold code at 1.023 million chips per second so that it repeats every millisecond. A chip is essentially the same thing as a bit and chips per second are the same as bits per second. The justification for coming up with this new term, chip, is that in some cases a sequence of bits is used as a type of modulation and contains no information.

Each satellite has its own C/A code so that it can be uniquely identified and received separately from the other satellites transmitting on the same frequency. The P-code is a 10.23 megachip per second PRN code that repeats only every week. When the "anti-spoofing" mode is on, as it is in normal operation, the P code is encrypted by the Y-code to produce the P(Y) code, which can only be decrypted by units with a valid decryption key. Both the C/A and P(Y) codes impart the precise time-of-day to the user.

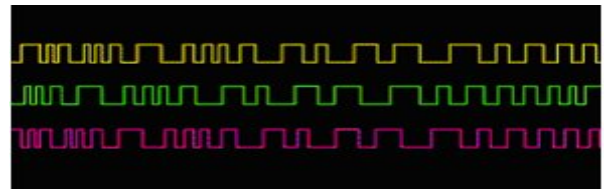
Fig.6: Unique pseudo random code of each satellite

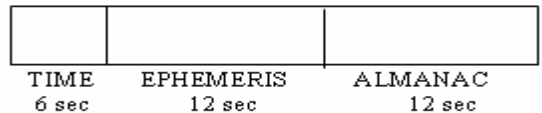
Fig7: Original gold code sent by satellite

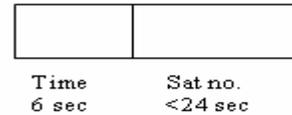
Fig 8: new gold code sent by satellite

*Other type of Gold Code*

From equation (i) we find, (tr-ti)c is the distance between satellite and receiver. If we consider while manufacturing we feed in the ephemeris and the almanac data of the satellites in the receiver. The receiver would now know at any instant of time where the satellite is

The signal sent would contain only the information about the satellite number and the time details. So the receiver would calculate the time taken by signal to reach the receiver and find the r. Then we apply triangulation with r = (tr-ti)c.

So we get the intersection of the spheres and find out the position rather quickly. This is the actual signal sent by the satellite. It contains the information about the time at which the message was sent in it's first six seconds of information. This information is necessary to calculate the position of the receiver. The next twelve seconds of information contains the rough health of the satellite and the orbit information of the satellite sending the signals. This data is known as ephemeris. The further twelve seconds contain the information about the orbits of all other satellites in the GPS grid. This data is known as almanac. This signal is of length 1500 bits and 30 seconds to be transmitted.

The proposed gold code is of very small length compared to the original gold code. It contains only the time segment of information and a small digital signal representing the satellite identity. The main modification here is that the orbital information of the satellite is fed inside the receiver while manufacturing. This is possible as the GPS satellite orbits rarely do change. The height of the satellites above the earth is such that it is not affected by the atmospheric effects. The receiver could thus calculate position without taking much time. This enhances the speed and thus the technology becomes better for application in AVL.

*Satellite frequencies*
L1 (1575.42 MHz): Mix of Navigation Message, coarse-acquisition (C/A) code and encrypted precision P(Y) code, plus the new L1C on future Block III satellites.
L2 (1227.60 MHz): P(Y) code, plus the new L2C code on the Block IIR-M and newer satellites.
L3 (1381.05 MHz): Used by the Nuclear Detonation (NUDET) Detection System Payload (NDS) to signal detection of nuclear detonations and other high-energy infrared events.

## VI. POSITION CALCULATION OF A VEHICLE

The cost was the biggest constraint in the mass aplication of the AVL. With the introduction of DGPS receivers the AVL can be took to the next level.
In general, vehicle navigation and AVL tasks can be broken down into three scales: 1) macroscale; 2) microscale; and 3) mesoscale.
*Macroscale:* The macroscale navigation usually performs the task of finding a particular path between two nodes in the network consisting of links (roadways) and nodes (e.g., intersections). This path is usually based on some optimality such as shortest distance or shortest duration. This scale navigation is used for commercial vehicles.
*Microscale:* The microscale level typically considers navigation at the vehicle level and is concerned with tasks such as lane keeping, as well as detecting and avoiding obstacles. At this level, there is no consideration of the ultimate or intermediate goal on the route. These tasks are generally carried out by the driver. This scale of navigation is used in anti collision device and unmanned vehicles.
*Mesoscale:* The mesoscale level, which is a level in between the microscale and macroscale, considers vehicle operation at the link level. A particular link may have a variety of features: multiple lanes, turn pockets, off-ramps, etc. From a navigation point of view, mesoscale route planning is generally concerned with vehicle maneuvers such as passing, pulling off to the side of the roadway, moving out of the way of emergency vehicles, merging in and out of special lanes (e.g., high-occupancy toll lanes), and choosing the correct lane to exit. A link-based planning algorithm may be concerned with when, where, and how lane changes are made with respect to a planned course change (e.g., turn and freeway exit) or the traffic situation.
This is the most effective navigational scale and was not being used till date due to cost constraint.

## VII. SYSTEM ARCHITECTURE OF LANE-LEVEL AVL SYSTEM
The AVL system consists of mainly four parts:
1) GPS satellites
2) GPRS communication
3) Base station
4) Rover or vehicle

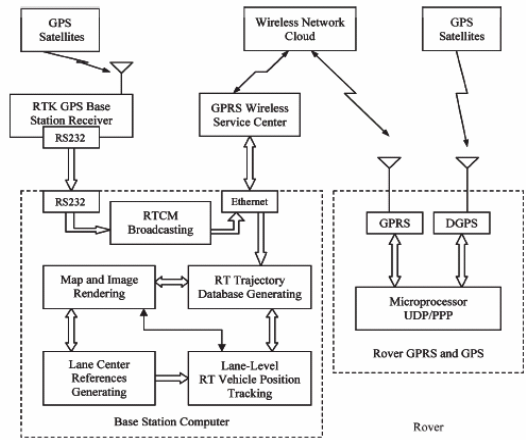

Fig.9:System architecture of lane –level AVL system

The functioning of the AVL (Fig. 9) can be divided in the following steps:
1) The process starts with the satellite sending the GPS signals which are received by the Rover and Base stations. The satellite signal is received by RTK GPS Base station receiver and the DGPS receiver at the Rover.
2) The receiver at the Base station calculates the error in the signal. The Base station position is predominantly fed inside the base. So the error in the signal due to multipath effect and clock bias can be calculated. The signal is also received by the Rover which calculates its position using triangulation.
3) The correction data from the base station is sent via GPRS to the Rover. The Rover utilizes this data to calculate its exact position and relays it back to the Base station using GPRS.
4) The position data received is sent to the Real Time (RT) Trajectory Database Generating section. Here the equation of the path of a vehicle is formed using the data sent in by the Rover. The equation is formed by considering the motion o the Rover in last few seconds and at that instant.
5) This trajectory is sent to the Map and Image Rendering Section. Here the trajectory equation is converted in pictorial form. This section is also helpful while creating a road map.
6) The trajectory picture is then sent to the Lane level Real Time (RT) vehicle position tracking section. This section does follow a process known as map matching. In this process the digital map of the road network and the trajectory picture of the Rover are matched. This gives the actual position of the Rover on the road network.

## VIII. LANE DETERMINATION ALGORITHM

The method of matching the trajectory points to the map network is called *map matching*. Different techniques exist for map matching, and they vary in complexity and appropriateness for certain applications. To formally define the problem, let us consider a vehicle moving along a system of streets $N$. At each of a finite number $T$ of discrete points in time, denoted by $\{0, 1, \ldots T\}$, the estimate $Pt$ of the position at time $t$ is available from the DGPS system, although the true position $P\ t$ at time $t$ is unknown. The goal is to match each estimated position $Pt$ to a lane in $N$. In other words, we need to determine what lane the vehicle is on at time $t$. Considering applications and data storage issue, the real lane system $N$ is represented by a digital network $N$, which precisely approximates the real lane system with the accuracy in accordance with lane-level navigation. The network $N$ consists of a set of curves in R, each of which is referred to as an *arc*. The arcs in $N$ are assumed to be piecewise linear. Fig. 10 illustrates an example of the approximation of an actual street by a piecewise linear curve. Because of this linear approximation, each arc $A \in N$ can be completely characterized by a finite sequence of points ($A0, A1, \ldots, An0$), each of which is in R. The beginning and ending points of this sequence, i.e., $A0$ and $An0$, respectively, will be defined as *nodes*, whereas the intermittent points ($A1, A2, \ldots, An0-1$) will be defined as *shape points*. The minimum number of shape points of a piecewise linear approximation curve is determined with reference to the curvature accuracy.

## IX. ROADWAY NETWORK DATA

The precise map of a place is highly important for the exact location and tracking of a vehicle. The digital map can be obtained by using probe vehicles. Vehicles fitted with DGPS can be used to get the image of the lanes on the map by using the map and image rendering module of the Base station. More will be the number of probes more will be the accuracy, but this is a cumbersome job. The second method being used now days is Human or Computer vision processing. It captures roadway feature data, lane number, probe occupied lane and lane end position.

*Map Matching Algorithm*

The trajectory as found out by the RT Trajectory Database Generating section has to be matched with the map of the lane system as already stored in the system. This task is completed by using an algorithm known as map matching algorithm. This process includes the conversion of the arc shaped lane network into piecewise linear poly lines. We denote the actual curve map as N and its digital conversion to poly line form as N'. The lanes in N are all represented by a poly line in N'. If the lane map data is stored in form of polygons rather than poly lines then the polygons have to be converted into poly lines. Let the polygon be defined by the points (L0,L1,……..Ln-1) then the corresponding points in the poly line would be (A0=(L0+L1)/2,A1=(L1+L2)/2,……….).

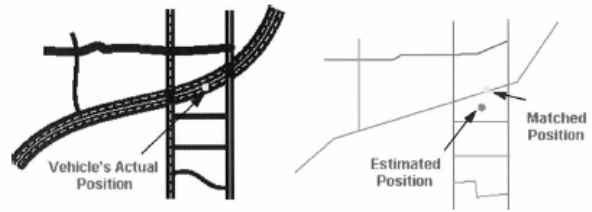

Fig10 : conversion of lane polygons to poly lines

The methods available for map matching are:1) point-to-point matching; 2) point to-curve matching; and 3) curve-to-curve matching. Out of these the most feasible and accurate is curve-to-curve-matching. The method is given as follows:

Let the latest section of the vehicle trajectory be represented as S= $\{P_{N-m+1}, \ldots, P_N\}$ a set of m points. The target is to match any point p with a lane polyline. The piecewise linear curve representing the trajectory of the vehicle is matched with the corresponding section of every lane polyline. The lane polylines chosen for this task are those which are within the distance $d$ (uncertainty of vehicle position estimate) of the point p. To find out the corresponding section of the lane polyline to the vehicle trajectory we find out the perpendicular distance between the lane polyline and the point p. This point is considered the first corresponding point on a given lane. The same process is followed for all the lanes within the feasible distance d. The next point on the lane is determined by moving forward along the lane a distance equal to the Euclidean distance $d(P_{N-m+1}, P_{N-m+2})$ between $P_{N-m+1}$ and $P_{N-m+2}$. These points along the lane are found until we have m points on the lane. The points $\{l_1, \ldots, l_m\}$ will then comprise a set, which represents a *corresponding segment* of the lane poly line. This set is denoted as C.

The process of matching one piecewise linear curve to another is accomplished through the integration of the distance between each corresponding segment of the two curves. The number of points in C and S are same. The distance between each corresponding line segment will then be integrated and summed on the curve segment to find the distance between curve C and the curve of vehicle trajectory. The corresponding line segment would be between points (PN-m+1,PN-m+2) on the trajectory polyline and ($l_1,l_2$) on the set C. The "distance" is defined here as the modified area enclosed by the trajectory curve segment $S$ and the curve segment $C$ of a candidate lane polyline. In determining the vehicle lane position, the "distances" of the vehicle trajectory curve $S$ to every candidate of lane polyline curves $Cs$ are calculated and compared. The candidate of lane polyline curve with the shortest "distance" is chosen as the lane on which the vehicle is running. The subset $S$ takes the final $m$ points of the vehicle trajectory section $T$ to match with the given lane polylines. As new vehicle position data are received, the result of the algorithm provides the vehicle's lane positioning information.

## X. ONBOARD VEHICLE ELECTRONICS

The Onboard Vehicle electronics consists of three primary components:
1) A microcontroller;
2) A GPRS wireless modem;
3) A differential capable GPS receiver (DGPS).

The microcontroller works as the main command system for the GPRS modem and the GPS receiver. It gives signals to the components on board to perform the task as per the input provided. That means when the GPS signals are received by the satellite the microcontroller commands the receiver to calculate its position and its velocity. It also manages the GPRS modem to establish a communication link with the server. If the GPRS signal is lost, the microcontroller initiates a reacquire process. During normal operation, the microcontroller processes interrupts from either the GPRS modem or the GPS receiver. If the signal comes from the GPS receiver, then the data are collected and sent out to the GPRS modem to send it to the base station. If an RTCM message arrives from the GPRS modem, then the corrections are processed and sent on to the GPS receiver to include them in the calculations. It works as a two way communication system between the GPRS modem and the GPS receiver. GPS data are sent at a rate of 1 Hz through the GPRS communication system. RTCM messages are acquired approximate once per 30 s and included in the position calculation data. The bandwidth requirements of this application are fairly low: well below the GPRS practical bandwidth of 20–50 kb/s. The GPRS modem module is capable of wide-area cellular data communication. The GPS module is extremely power efficient and capable of differential operation. So it can easily be fitted inside vehicles at low cost. Moreover the GPRS communication is a fast and reliable method.

## XI. CONCLUSION

The overall goal of this project was to develop a low cost Real Time AVL system that can report lane-level position information of vehicles in real time. The developed system has met this goal and has been demonstrated successfully in real-time road tests. The developed onboard electronic unit can consistently position the vehicle in real time at the lane accuracy level. As a major part of this project; two methods have been developed to construct lane-level roadway network data. One method is based on an *in situ* driving method, and the other uses high-resolution aerial images. Overall, the aerial image technique is quicker and more accurate for constructing the lane data structures; however it requires that high-resolution imagery be available. For areas without aerial images, the ground-based *in situ* driving technique is also effective but requires that all the roadways be driven and coded to create the lane data structures. Other conclusions from this project can be made.
1) The accuracy and response time are sufficient for most applications, although the data from the vehicle are received in real-time operation at the base station with an approximate delay of 5–10 s. The latency is primarily due to delays along the wireless channels of the GPRS modems and other delays that are associated with the Internet.
2) The developed lane-matching technique works fairly well in real-time operation with the modified curve-to-curve matching technique using a trajectory generated from the AVL-equipped vehicle. However, it can be improved by using prior knowledge of vehicle motion constraints. For example, during a lane change, a characteristic "S"- shape curve is typically followed by most vehicles. Such knowledge can be exploited to improve lane matching during a lane-change event.

## REFERENCES

[1] Jie Du and Matthew J. Barth **"**Next-Generation Automated Vehicle Location Systems: Positioning at the Lane Level

[2] I. Chabini and S. Lan, "Adaptation of A∗ algorithm for the computation of fastest path in deterministic discrete-time dynamic networks," *IEEE Trans. Intell. Transp. Syst.*, vol. 3, no. 1, pp. 60–74, Mar. 2002.

[3] T. R. Connolly and J. K. Hedrick, "Longitudinal transition maneuvers in an automated highway system," *Trans. ASME, J. Dyn. Syst. Meas. Control*, vol. 121, no. 3, pp. 471–478, Sep. 1999.

[4] C. Hatipoglu, U. Ozguner, and K. A. Redmill, "Automated lane change controller design," *IEEE Trans. Intell. Transp. Syst.*, vol. 4, no. 1, pp. 13–22, Mar. 2003.

[5] G. Lu and M. Tomizuka, "Vehicle lateral control with combined use of a laser scanning radar sensor and rear magnetometers," in *Proc. Amer.Control Conf.*, Danvers, MA, 2002, vol. 5, pp. 3702–3707.

[6] J. K. Hedrick, M. Tomizuka, and P. Varaiya, "Control issues in automated highway systems," *IEEE Control Syst. Mag.*, vol. 14, no. 6, pp. 21–32,Dec. 1994.

[7] R. Horowitz and P. Varaiya, "Control design of an automated highway system," *Proc. IEEE*, vol. 88, no. 7, pp. 913–925, Jul. 2000.

[8] J. Farrell and M. Barth, *The Global Positioning System and Inertial Navigation: Theory and Practice*. New York: McGraw-Hill, Jan. 1999.

[9] J. Farrell and T. Givargis, "Differential GPS reference station algorithm: Design and analysis," *IEEE Trans. Control Syst. Technol.*, vol. 8, no. 3, pp. 519–531, May 2000.

[10] H. Blomenhofer *et al.*, "Development of a real-time DGPS system in the centimeter range," in *Proc. IEEE Position Location Navigation Symp.*, Las Vegas, NV, 1994, pp. 532–539

[11] C. Kee and B. Parkinson, "Wide area differential GPS as a future navigation system in the U.S.," in *Proc. IEEE Position Location Navigation Symp.*, Las Vegas, NV, 1994, pp. 788–795.

[12] "Navstar GPS Interface Control Document," El Segundo, CA, Rep. ICDGPS-e 200B-PR. [Online].